\documentclass[twocolumn,prc,floatfix,showkeys,showpacs,preprintnumbers,amsmath,amssymb]{revtex4}

\usepackage{graphicx}
\usepackage{dcolumn}
\usepackage{bm}

\begin{document}


\title{Origin of the neutron skin thickness of $^{208}$Pb
in nuclear mean-field models}

\author{M. Centelles\textsuperscript{1}}
 \email{mariocentelles@ub.edu}
\author{X. Roca-Maza\textsuperscript{1}}
 \email{roca@ecm.ub.es}
\author{X. Vi\~nas\textsuperscript{1}}
 \email{xavier@ecm.ub.es}
\author{M. Warda\textsuperscript{1,2}}
 \email{warda@kft.umcs.lublin.pl}

\affiliation{\textsuperscript{1}Departament d'Estructura i Constituents de la Mat\`eria
and Institut de Ci\`encies del Cosmos,
Facultat de F\'{\i}sica, Universitat de Barcelona,
Diagonal {\sl 647}, {\sl 08028} Barcelona, Spain\\
\textsuperscript{2}Katedra Fizyki Teoretycznej, Uniwersytet Marii Curie--Sk\l odowskiej,
        ul. Radziszewskiego 10, 20-031 Lublin, Poland}

\date{\today}

\begin{abstract}
We study whether the neutron skin thickness $\Delta r_{np}$ of
$^{208}$Pb originates from the bulk or from the surface of the nucleon
density distributions, according to the mean-field models of nuclear
structure, and find that it depends on the stiffness of the nuclear
symmetry energy. The bulk contribution to $\Delta r_{np}$ arises from
an extended sharp radius of neutrons, whereas the surface contribution
arises from different widths of the neutron and proton surfaces.
Nuclear models where the symmetry energy is stiff, as typical
relativistic models, predict a bulk contribution in $\Delta r_{np}$ of
$^{208}$Pb about twice as large as the surface contribution. In
contrast, models with a soft symmetry energy like common
nonrelativistic models predict that $\Delta r_{np}$ of $^{208}$Pb is
divided similarly into bulk and surface parts. Indeed, if the symmetry
energy is supersoft, the surface contribution becomes dominant. We note
that the linear correlation of $\Delta r_{np}$ of $^{208}$Pb with the
density derivative of the nuclear symmetry energy arises from the bulk
part of $\Delta r_{np}$. We also note that most models predict a
mixed-type (between halo and skin) neutron distribution for
$^{208}$Pb. Although the halo-type limit is actually found in the
models with a supersoft symmetry energy, the skin-type limit is not
supported by any mean-field model. Finally, we compute
parity-violating electron scattering in the conditions of the $^{208}$Pb
parity radius experiment (PREX) and obtain a pocket formula for the
parity-violating asymmetry in terms of the parameters that
characterize the shape of the $^{208}$Pb nucleon densities.
\end{abstract}


\maketitle


\section{Introduction}


The study of the size and shape of the density distributions of
protons and neutrons in nuclei is a classic, yet always contemporary
area of nuclear physics. The proton densities of a host of nuclei are
known quite well from the accurate nuclear charge densities measured
in experiments involving the electromagnetic interaction
\cite{fricke95}, like elastic electron scattering. In contrast,
the neutron densities have been probed in fewer nuclei and are
generally much less certain.

The neutron distribution of $^{208}$Pb, and its rms radius in
particular, is nowadays attracting significant interest in both
experiment and theory. Indeed, the neutron skin thickness, i.e.,
the neutron-proton rms radius difference
\begin{equation}
\label{skin}
\Delta r_{np}= \langle r^2 \rangle_n^{1/2} 
- \langle r^2 \rangle_p^{1/2} ,
\end{equation}
of this nucleus has close ties with the density-dependent nuclear
symmetry energy and with the equation of state of neutron-rich matter.
In nuclear models, $\Delta r_{np}$ of $^{208}$Pb displays nearly
linear correlations with the slope of the equation of state of neutron
matter \cite{bro00,typ01,cen02}, with the density derivative $L$ of the
symmetry energy \cite{fur02,dan03,bal04,ava07,cen09,cen09b,vid09}, and
with the surface-symmetry energy of the finite nucleus \cite{cen09}.
At first sight, it may seem intriguing that a property of the mean
position of the surface of the nucleon densities ($\Delta r_{np}$) is
correlated with a purely bulk property of infinite nuclear matter
($L$). However, we have to keep in mind that $\Delta r_{np}$ depends
on the surface-symmetry energy. This quantity reduces the bulk-symmetry 
energy due to the finite size of the nucleus. Assuming a
local density approximation, we can correlate the surface-symmetry
energy with the density slope $L$, which determines the departure
of the symmetry energy from the bulk value. The correlation of $\Delta
r_{np}$ with $L$ then follows. Actually, these correlations can be
derived almost analytically starting from the droplet model (DM) of
Myers and \'{S}wi\c{a}tecki \cite{MS,MSskin} as we showed in
Refs.\ \cite{cen09,cen09b}.
By reason of its close connections with the nuclear
symmetry energy, knowing accurately $\Delta r_{np}$ of
$^{208}$Pb can have important implications in diverse problems of
nuclear structure and of heavy-ion reactions, in studies of atomic
parity violation, as well as in the description of neutron stars and
in other areas of nuclear astrophysics (see, e.g., Refs.\
\cite{hor01,die03,sil05,ste05a,lat07,dhiman07,samaddar07,kli07,li08, 
xu09,sun09,khoa09,cen10,carbone10,chen10}). Since the charge radius of
$^{208}$Pb has been measured with extreme accuracy ($r_{ch}=
5.5013(7)$ fm \cite{fricke95}), the neutron rms radius of $^{208}$Pb
is the principal unknown piece of the puzzle.

The lead parity radius experiment (PREX) \cite{prex1} is a challenging
experimental effort that aims to determine $\langle r^2
\rangle_n^{1/2}$ of $^{208}$Pb almost model independently and to 1\%
accuracy by parity-violating electron scattering \cite{prex1,prex2}.
This purely electroweak experiment has been run at the Jefferson Lab
very recently, although results are not yet available. The 
 parity-violating electron scattering is useful to measure neutron
densities because in the low-momentum transfer regime the $Z^0$
boson couples mainly to neutrons. For protons, this coupling is highly
suppressed because of the value of the Weinberg angle. Therefore, from
parity-violating electron scattering one can obtain the weak charge
form factor and the closely related neutron form factor. From these
data, the neutron rms radius can in principle be deduced \cite{prex2}.
This way of proceeding is similar to how the charge density is
obtained from unpolarized electron scattering data \cite{prex2}. The
electroweak experiments get rid of the complexities of the hadronic
interactions and the reaction mechanism does not have to be modeled.
Thus, the analysis of the data can be both clean and model
independent. There may be a certain model dependence, in the end, in
having to use some neutron density shape to extract the neutron rms
radius from the parity-violating asymmetry measured at a finite
momentum transfer.

To date, the existing constraints on neutron radii, skins, and
neutron distributions of nuclei have mostly used strongly interacting
hadronic probes. Unfortunately, the measurements of neutron
distributions with hadronic probes are bound to have some model
dependence because of the uncertainties associated with the strong
force. Among the more frequent experimental techniques we may quote
nucleon elastic scattering \cite{kar02,cla03}, the inelastic
excitation of the giant dipole and spin-dipole resonances
\cite{kra99,kra04}, and experiments in exotic atoms
\cite{trz01,jas04,fried03,fried05}. Recent studies indicate that the
pygmy dipole resonance may be another helpful tool to constrain
neutron skins \cite{kli07,carbone10}.

The extraction of neutron radii and neutron skins from the experiment
is intertwined with the dependence of these quantities on the shape of
the neutron distribution \cite{trz01,jas04,fried03,fried05,don09}. The
data typically do not indicate unambiguosly, by themselves, if the
difference between the peripheral neutron and proton densities that
gives rise to the neutron skin is caused by an extended bulk radius of
the neutron density, by a modification of the width of the surface, or
by some combination of both effects. In the present work we look for
theoretical indications on this problem and study whether the origin
of the neutron skin thickness of $^{208}$Pb comes from the bulk or
from the surface of the nucleon densities according to the mean-field
models of nuclear structure. The answer turns out to be connected with
the density dependence of the nuclear symmetry energy in the theory.

We described in Ref.\ \cite{warda10} a procedure to discern bulk
and surface contributions in the neutron skin thickness of nuclei. It
can be applied to both theoretical and experimental nucleon densities
as it only requires to know the equivalent sharp radius and surface
width of these densities, which one can obtain by fitting the actual
densities with two-parameter Fermi (2pF) distributions. The 2pF shape
is commonly used to characterize nuclear densities and nuclear
potentials in both theoretical and experimental analyses. 
The doubly magic number of protons and neutrons in $^{208}$Pb ensures
that deformations do not influence the results and spherical density
distributions describe the nuclear surface very well. 
We perform our calculations with several representative effective
nuclear forces, namely, nonrelativistic interactions of the Skyrme and
Gogny type and relativistic mean-field (RMF) interactions. The free
parameters and coupling constants of these nuclear interactions have
usually been adjusted to describe data that are well known empirically
such as binding energies, charge radii, single-particle properties,
and several features of the nuclear equation of state. However, the
same interactions predict widely different results for the size of
neutron skin of $^{208}$Pb and, as we will see, for its bulk or
surface nature. We also study the halo or skin character
\cite{trz01,jas04,fried03,fried05,don09} of the nucleon densities of
$^{208}$Pb in mean-field models. Finally, we perform calculations
of parity-violating electron scattering on $^{208}$Pb. We show
that if 2pF nucleon densities are assumed, the parity-violating
asymmetry as predicted by mean-field models can be approximated by a
simple and analytical expression in terms of the central radius and
surface width of the neutron and proton density profiles. This
suggests that an experiment such as PREX could allow to obtain some
information about the neutron density profile of the $^{208}$Pb
nucleus in addition to its neutron rms radius.

The rest of the article proceeds as follows. In Sec.\ \ref{formalism},
we summarize the formalism to decompose the neutron skin thickness into
bulk and surface components. The results obtained in the nuclear
mean-field models are presented and discussed in Sec.\ \ref{results}.
A summary and the conclusions are given in Sec.\ \ref{summary}.


\section{Formalism}
\label{formalism}


The analysis of bulk and surface contributions to the neutron skin
thickness of a nucleus requires proper definitions of these quantities
based on nuclear density distributions. We presented in Ref.\
\cite{warda10} such a study, and we summarize only its basic points here.

One can characterize the size of a nuclear density distribution
$\rho(r)$ through several definitions of radii, and each definition
may be more useful for a specific purpose (see Ref.\ \cite{has88} for
a thorough review). Among the most common radii, we have the {\it
central radius}~$C$:
\begin{equation} 
\label{c}
C= \frac{1}{\rho(0)} \int_0^{\infty} \rho(r) dr \,;
\end{equation} 
the {\it equivalent sharp radius}~$R$:
\begin{equation} 
\label{r}
\frac 43 \pi R^3 \rho({\rm bulk}) =
4\pi \int_0^{\infty} \rho(r) r^2 dr ,
\end{equation} 
i.e., the radius of a uniform sharp distribution whose density equals
the bulk value of the actual density and has the same number of
particles; and the {\it equivalent rms radius}~$Q$:
\begin{equation} 
\label{q}
\frac 35 \, Q^2= \langle r^2 \rangle ,
\end{equation} 
which describes a uniform sharp distribution with the same rms radius
as the given density. These three radii are related by expansion
formulas \cite{has88}:
\begin{equation} 
\label{qr}
Q = R \Big(1+\frac 52\frac {b^2}{R^2} + \ldots \Big) ,
\quad
C = R \Big(1-\frac {b^2}{R^2} + \ldots \Big) .
\end{equation} 
Here, $b$ is the {\it surface width} of the density profile:
\begin{equation} 
\label{b}
b^2= - \frac{1}{\rho(0)} 
\int_0^{\infty} (r-C)^2 \frac{d \rho(r)}{dr} dr ,
\end{equation} 
which provides a measure of the extent of the surface of the density.
Relations (\ref{qr}) usually converge quickly because $b/R$ is
small in nuclei, especially in heavy-mass systems.
Nuclear density distributions have oscillations in the
inner bulk region and a meaningful average is needed to determine the
density values $\rho(0)$ and $\rho({\rm bulk})$ appearing in the above
equations. This can be achieved by matching the original density with
a 2pF distribution:
\begin{equation} 
\label{2pf}
\rho(r) = 
\frac{\rho_0}{1+ \exp{ [(r-C)/a] }} .
\end{equation} 
In 2pF functions the bulk density value corresponds very closely to
the central density, and the latter coincides to high accuracy with
the $\rho_0$ parameter if $\exp{(-C/a)}$ is negligible. The surface
width $b$ and the diffuseness parameter $a$ of a 2pF function are
related by $b= (\pi/ \sqrt{3}) a$.

\begin{figure}
\includegraphics[width=0.98\columnwidth,clip=true]
{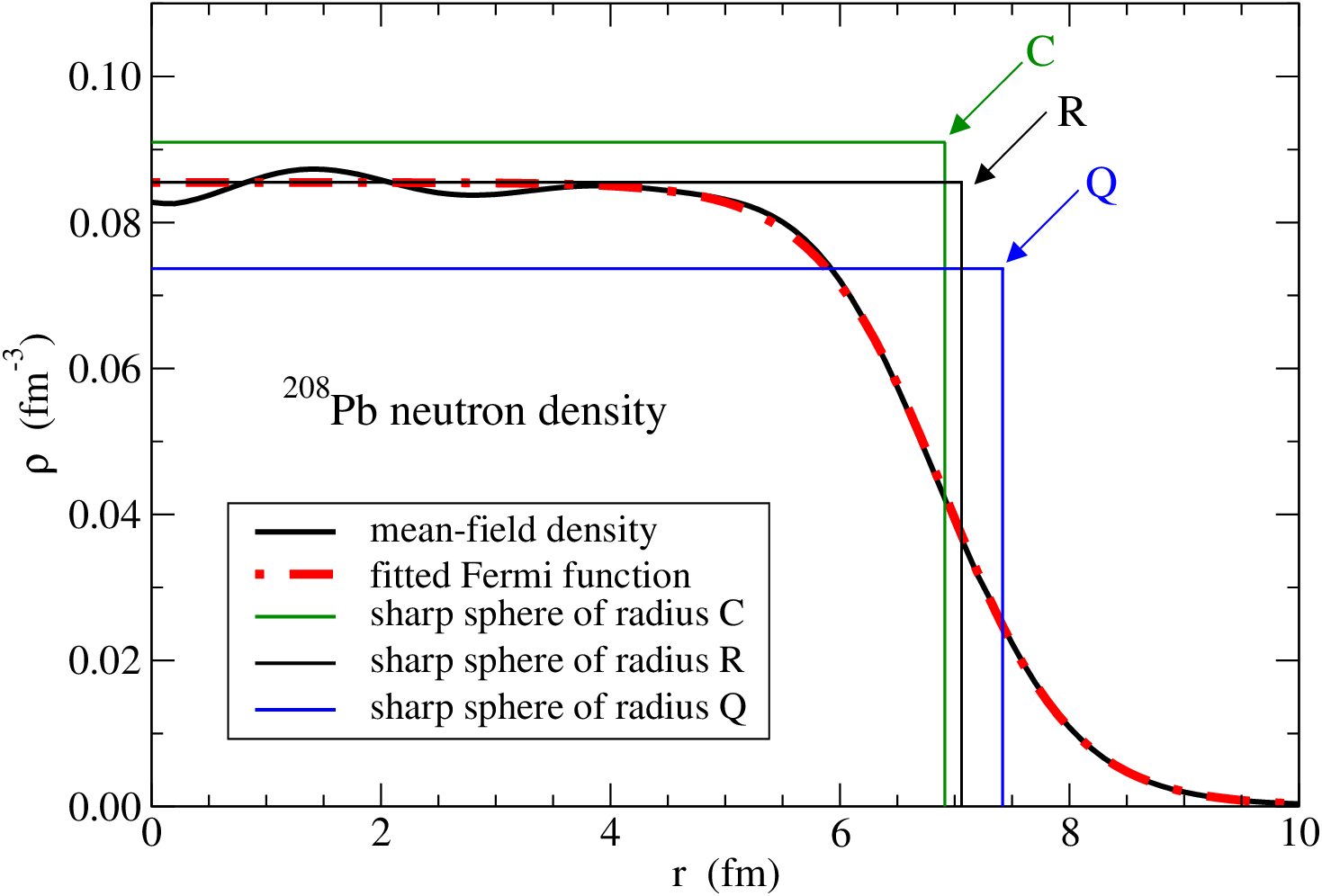}
\caption{\label{radii} 
(Color online)
Comparison of sharp density profiles having radii $C$, $R$, and $Q$
with the mean-field and 2pF density distributions for the neutron
density of $^{208}$Pb. The RMF interaction NL3 has been used
in the mean-field calculation.}
\end{figure}

As discussed in Ref.\ \cite{has88}, the equivalent sharp radius $R$ is
the quantity of basic geometric importance of the $C$, $Q$, and $R$
radii. This is because a sharp distribution of radius $R$ has the same
volume integral as the density of the finite nucleus and differs from
it only in the surface region. We illustrate this fact in
Fig.~\ref{radii} using as example the neutron density of $^{208}$Pb of
a mean-field calculation. We can see that the mean-field density is
clearly overestimated in the whole nuclear interior by a sharp sphere
of radius $C$. The equivalent rms radius $Q$ fails also, by
underestimating it. Only the equivalent sharp radius $R$ is able to
reproduce properly the bulk part of the original density profile of
the nucleus. Therefore, $R$ appears as the suitable radius to describe
the size of the bulk of the nucleus.

As the neutron skin thickness (\ref{skin}) is defined through rms
radii, it can be expressed with $Q$:
\begin{equation}
\label{r0}
\Delta r_{np}=\sqrt{\frac{3}{5}} \left(Q_n-Q_p\right) .
\end{equation}
Recalling from (\ref{qr}) that $Q\simeq R + \frac 52 (b^2/R)$,
we have a natural distinction in
$\Delta r_{np}$ between bulk ($\propto R_n-R_p$) and surface
contributions. That is to say,
\begin{equation}
\label{rtot}
\Delta r_{np} = \Delta r_{np}^{\rm bulk} + \Delta r_{np}^{\rm surf} ,
\end{equation}
with
\begin{equation}
\label{rb}
\Delta r_{np}^{\rm bulk} = \sqrt{\frac{3}{5}}\left(R_n-R_p\right)
\end{equation}
independent of surface properties, and
\begin{equation}
\label{rs}
\Delta r_{np}^{\rm surf} = \sqrt{\frac{3}{5}} \, \frac{5}{2}
   \Big(\frac{b_n^2}{R_n}-\frac{b_p^2}{R_p}\Big)
\end{equation}
of surface origin.
The nucleus may develop a neutron skin by separation of the bulk radii
$R$ of neutrons and protons or by modification of the width $b$ of the
surfaces of the neutron and proton densities. In the general case,
both effects are expected to contribute. 
We note that Eq.\ (\ref{rs}) coincides with the expression of the
surface width contribution to the neutron skin thickness provided by
the DM of Myers and \'{S}wi\c{a}tecki \cite{MS,MSskin} if we set 
in Eq.\ (\ref{rs}) $R_n= R_p= r_0 A^{1/3}$.

The next-order correction to Eq.\ (\ref{rs}) can be easily evaluated
for 2pF distributions (cf.\ Ref.\ \cite{has88} for the higher-order
corrections to the expansions (\ref{qr})) and gives
\begin{equation}
\label{rscorr}
\Delta r_{np}^{\rm surf,corr} = - \sqrt{\frac{3}{5}} \, \frac{5}{2} \,
\frac{21}{20} \Big(\frac{b_n^4}{R_n^3}-\frac{b_p^4}{R_p^3}\Big) .
\end{equation}
This quantity is usually very small---indeed, we neglected it in
\cite{warda10}. In the case of $^{208}$Pb, we have found in all
calculations with mean-field models that it is between
$-0.0025$ fm and $-0.004$ fm, and thus can be obviated for most purposes.
But because in the present work we deal with some detailed
comparisons among the models, we have included (\ref{rscorr}) in the
numerical values shown for the surface contribution $\Delta
r_{np}^{\rm surf}$ in the later sections.

It is to be mentioned that there is no universal method to do the
parametrization of the neutron and proton densities with 2pF
functions. A popular prescription is to use a $\chi^2$ minimization of
the differences between the density to be reproduced and the 2pF
profile, or of the differences between their logarithms. These methods
may somewhat depend on conditions given during minimization (number of
mesh points, limits, etc.). As in \cite{warda10}, we have preferred to
extract the parameters of the 2pF profiles by imposing that they
reproduce the same quadratic $\langle r^2 \rangle$ and quartic
$\langle r^4 \rangle$ moments of the self-consistent mean-field
densities, and the same number of nucleons. These conditions allow us
to determine in a unique way the equivalent 2pF densities and pay
attention to a good reproduction of the surface region of the original
density because the local distributions of the quantities $r^2
\rho(r)$ and $r^4 \rho(r)$ are peaked at the peripheral region of the
nucleus. An example of this type of fit is displayed in
Fig.~\ref{radii} by the dash-dotted line. It can be seen that the
equivalent 2pF distribution nicely averages the quantal oscillations
at the interior and reproduces accurately the behavior of the
mean-field density at the surface.


\section{Results}
\label{results}


\subsection{Survey of model predictions and of data for
$\Delta r_{np}$ of $^{208}$Pb}

We calculate our results with nonrelativistic models of Skyrme
type (SGII, Ska, SkM*, Sk-T4, Sk-T6, Sk-Rs, SkMP, SkSM*, MSk7, SLy4,
HFB-8, HFB-17) and of Gogny type (D1S, D1N), as well as with several
relativistic models (NL1, NL-Z, NL-SH NL-RA1, NL3, TM1, NLC, G2,
FSUGold, DD-ME2, NL3*). The original references to the different
interactions can be found in the papers \cite{xu09,HFB-17} for the
Skyrme models, \cite{cha08} for the Gogny models, and
\cite{patra02b,sulaksono07,FSUG,DDME2,NL3S} and Ref.\ [19] in
\cite{patra02b} for the RMF models. It may be mentioned that the
recent force HFB-17 \cite{HFB-17} achieves the lowest rms deviation
with respect to experimental nuclear masses found to date in a
mean-field approach. 

As is well known, nonrelativistic and relativistic models differ in
the stiffness of the symmetry energy. Note that by soft or stiff
symmetry energy we mean that the symmetry energy increases slowly or
rapidly as a function of the nuclear density around the saturation
point. Of course, the soft or stiff character can depend on the
explored density region; for example, it is possible that a symmetry
energy that is soft at nuclear densities becomes stiff at much higher
densities \cite{HFBXII}, or that a model with a stiff symmetry energy
at normal density has a smaller symmetry energy at low densities
\cite{todd03}. The density dependence of the nuclear symmetry energy
$c_{\rm sym}(\rho)$ around saturation is frequently parametrized
through the slope $L$ of $c_{\rm sym}(\rho)$ at the saturation density: 
\begin{equation}
L = \left. 3\rho_0\frac{\partial c_{\rm sym}(\rho)}{\partial \rho}
\right|_{\rho_0} .
\label{Lsym}
\end{equation}
The pressure of pure neutron matter is directly proportional to $L$
\cite{piek09} and thus the $L$ value has important implications for
both neutron-rich nuclei and neutron stars. The symmetry energy of the
Skyrme and Gogny forces analyzed in this work displays, as usual in
the nonrelativistic models, from the very soft to the moderately stiff
density dependence at nuclear densities (see Table~\ref{TABLE1} for
the $L$ parameter of the models). On the contrary, the majority of the
relativistic parameter sets have a stiff or very stiff symmetry energy
around saturation. The exception to the last statement in our case are
the covariant parameter sets FSUGold and DD-ME2 that have a milder
symmetry energy than the typical RMF models. FSUGold achieves this
through an isoscalar-isovector nonlinear meson coupling \cite{FSUG}
and DD-ME2 because of having density-dependent meson-exchange
couplings \cite{DDME2}.

In Table~\ref{TABLE1} we display the neutron skin thickness of
$^{208}$Pb obtained from the self-consistent densities of the various
interactions (denoted as $\Delta r_{np}^{\rm s.c.}$). It is
evident that the nuclear mean-field models predict a large window of
values for this quantity. The nonrelativistic models with softer
symmetry energies point toward a range of about 0.1--0.17 fm. Most of
the relativistic models, having a stiff symmetry energy, point toward
larger neutron skins of 0.25--0.3~fm. In between, the relativistic
models DD-ME2 and FSUGold predict a result close to 0.2~fm and the
Skyrme interactions that have relatively stiffer symmetry energies
fill in the range between 0.2 and 0.25~fm.

\begin{table}[t]
\caption{Neutron skin thickness in $^{208}$Pb calculated with the
self-consistent densities of several nuclear mean-field models
($\Delta r_{np}^{\rm s.c.}$) and its partition into bulk and surface
contributions defined in Sec.\ \ref{formalism}, as well as the
relative weight of these bulk and surface parts. The models have been
set out according to increasing $\Delta r_{np}^{\rm s.c.}$. The density
slope $L$ of the symmetry energy of the models is also listed.
In order to help distinguish relativistic and nonrelativistic models,
we have preceded the relativistic ones with an r in this table.}
\begin{ruledtabular}
\begin{tabular}{lccccccc}
Model & $\Delta r_{np}^{\rm s.c.}$ & 
$\Delta r_{np}^{\rm bulk}$ & $\Delta r_{np}^{\rm surf}$ & 
bulk & surf & $L$ \\
        & (fm)  & (fm)  & (fm)  & \% & \% & (MeV) \\
\hline
\; HFB-8   & 0.115 & 0.031 & 0.084 & 27 & 73 & \ 14.8 \\
\; MSk7    & 0.116 & 0.030 & 0.086 & 26 & 74 & \ \ 9.4\\
\; D1S     & 0.135 & 0.062 & 0.073 & 46 & 54 & \ 22.4 \\
\; SGII    & 0.136 & 0.065 & 0.071 & 48 & 52 & \ 37.6 \\
\; D1N     & 0.142 & 0.070 & 0.072 & 49 & 51 & \ 31.9 \\
\; Sk-T6   & 0.151 & 0.067 & 0.084 & 44 & 56 & \ 30.9 \\
\; HFB-17  & 0.151 & 0.067 & 0.084 & 44 & 56 & \ 36.3 \\
\; SLy4    & 0.161 & 0.086 & 0.075 & 53 & 47 & \ 46.0 \\
\; SkM*    & 0.170 & 0.093 & 0.077 & 55 & 45 & \ 45.8 \\
r\! DD-ME2  & 0.193 & 0.098 & 0.095 & 51 & 49 & \ 51.3 \\
\; SkSM*   & 0.197 & 0.116 & 0.082 & 58 & 42 & \ 65.5 \\
\; SkMP    & 0.197 & 0.123 & 0.074 & 62 & 38 & \ 70.3 \\
r\! FSUGold & 0.207 & 0.105 & 0.102 & 51 & 49 & \ 60.5 \\
\; Ska     & 0.211 & 0.140 & 0.071 & 66 & 34 & \ 74.6 \\
\; Sk-Rs   & 0.215 & 0.146 & 0.069 & 68 & 32 & \ 85.7 \\
\; Sk-T4   & 0.248 & 0.163 & 0.085 & 66 & 34 & \ 94.1 \\
r\! G2      & 0.257 & 0.171 & 0.086 & 66 & 34 &  100.7 \\
r\! NLC     & 0.263 & 0.174 & 0.089 & 66 & 34 &  108.0 \\
r\! NL-SH   & 0.266 & 0.169 & 0.097 & 64 & 36 &  113.6 \\
r\! TM1     & 0.271 & 0.172 & 0.098 & 64 & 36 &  110.8 \\
r\! NL-RA1  & 0.274 & 0.179 & 0.095 & 65 & 35 &  115.4 \\
r\! NL3     & 0.280 & 0.185 & 0.095 & 66 & 34 &  118.5 \\
r\! NL3*    & 0.288 & 0.191 & 0.097 & 66 & 34 &  122.6 \\
r\! NL-Z    & 0.307 & 0.209 & 0.098 & 68 & 32 &  133.3 \\
r\! NL1     & 0.321 & 0.216 & 0.105 & 67 & 33 &  140.1
\end{tabular}
\end{ruledtabular}
\label{TABLE1}
\end{table}

Before proceeding, we would like to briefly survey some of the recent
results deduced for $\Delta r_{np}$ in $^{208}$Pb from experiment. For
example, the recent analysis in Ref.\ \cite{klo07} of the data
measured in the antiprotonic $^{208}$Pb atom \cite{trz01,jas04} gives
$\Delta r_{np}= 0.16 \pm(0.02)_{\rm stat} \pm(0.04)_{\rm syst}$ fm,
including statistical and systematic errors. Another recent study
\cite{bro07} of the antiprotonic data for the same nucleus leads to
$\Delta r_{np}= 0.20 \pm(0.04)_{\rm exp} \pm(0.05)_{\rm th}$ fm, where
the theoretical error is suggested from comparison of the models
with the experimental charge density.
These determinations are in consonance with the {\em average} value of
the hadron scattering data for the neutron skin thickness of
$^{208}$Pb, namely, $\Delta r_{np} \sim 0.165 \pm 0.025$ fm (taken
from the compilation of hadron scattering data in Fig.~3 of Ref.\
\cite{jas04}). We may also mention that the constraints on the nuclear
symmetry energy derived from isospin diffusion in heavy ion collisions
of neutron-rich nuclei suggest $\Delta r_{np}= 0.22\pm 0.04$ fm in
$^{208}$Pb \cite{che05}. Following Ref.\ \cite{ste05}, the same type
of constraints exclude $\Delta r_{np}$ values in $^{208}$Pb less than
0.15 fm. A recent prediction based on measurements of
the pygmy dipole resonance in $^{68}$Ni and $^{132}$Sn gives $\Delta
r_{np}= 0.194 \pm 0.024$ fm in $^{208}$Pb \cite{carbone10}. 
Finally, we quote the new value $\Delta r_{np}=
0.211^{+0.054}_{-0.063}$ fm determined in \cite{zenihiro10} from
proton elastic scattering. Thus, in
view of the empirical information for the central value of $\Delta
r_{np}$ and in view of the $\Delta r_{np}^{\rm s.c.}$ values predicted
by the theoretical models in Table~\ref{TABLE1}, it may be said that
those interactions with a soft (but not very soft) symmetry energy,
for example, HFB-17, SLy4, SkM*, DD-ME2, or FSUGold, agree better with
the determinations from experiment. Nevertheless, the uncertainties in
the available information for $\Delta r_{np}$ are rather large and one
cannot rule out the predictions by other interactions. If the PREX
experiment \cite{prex1,prex2} achieves the purported goal of
accurately measuring the neutron rms radius of $^{208}$Pb, it will
allow to pin down more strictly the constraints on the neutron
skin thickness of the mean-field models.

\subsection{Bulk and surface contributions to $\Delta r_{np}$ of
$^{208}$Pb in nuclear models and the symmetry energy}
\label{bulksurf}

We next discuss the results for the division of the neutron skin
thickness of $^{208}$Pb into bulk ($\Delta r_{np}^{\rm bulk}$) and
surface ($\Delta r_{np}^{\rm surf}$) contributions in the nuclear
mean-field models, following Sec.\ \ref{formalism}. We display this
information in Table~\ref{TABLE1}. It may be noticed that the value of
$\Delta r_{np}^{\rm bulk}$ plus $\Delta r_{np}^{\rm surf}$ (quantities
obtained from Eqs.\ (\ref{rb})--(\ref{rscorr})) agrees excellently
with $\Delta r_{np}^{\rm s.c.}$ (neutron skin thickness obtained from
the self-consistent densities). One finds that the bulk contribution
$\Delta r_{np}^{\rm bulk}$ to the neutron skin of $^{208}$Pb varies in
a window from about 0.03 fm to 0.22 fm. The surface contribution
$\Delta r_{np}^{\rm surf}$ is comprised approximately between 0.07 fm
and 0.085 fm in the nonrelativistic forces, and between 0.085 fm and
0.105 fm in the relativistic ones. Thus, whereas the bulk
contribution to the neutron skin thickness of $^{208}$Pb changes
largely among the different mean-field models, the surface
contribution remains confined to a narrower band of values.

Table~\ref{TABLE1} shows that the size of the neutron skin thickness
of $^{208}$Pb is divided into bulk and surface contributions in almost
equal parts in the nuclear interactions that have soft symmetry
energies (say, $L\sim20$--60). This is the case of multiple
nonrelativistic interactions and of the covariant DD-ME2 and FSUGold
parameter sets. When the symmetry energy becomes softer, the bulk part
tends to be smaller. Indeed, we see that in the models that
have a very soft symmetry energy ($L\lesssim20$), which we may call
``supersoft'' \cite{wen09}, the surface contribution takes over and it
is responsible for the largest part ($\sim75\%$) of $\Delta r_{np}$ of
$^{208}$Pb. At variance with this situation, in the models with
stiffer symmetry energies ($L\gtrsim75$) about two thirds of $\Delta
r_{np}$ of $^{208}$Pb come from the bulk contribution, as seen in the
Skyrme forces of stiffer symmetry energy and in all of the
relativistic forces that have a conventional isovector channel (G2,
TM1, NL3, etc.). We therefore note that in a heavy neutron-rich
nucleus with a sizable neutron skin such as $^{208}$Pb, the nuclear
interactions with a soft symmetry energy predict that the contribution
to $\Delta r_{np}$ produced by differing widths of the surfaces of the
neutron and proton densities ($b_n \ne b_p$) is similar to, or even
larger than, the effect from differing extensions of the bulk of the
nucleon densities ($R_n \ne R_p$). On the contrary, the nuclear
interactions with a stiff symmetry energy favor a dominant bulk nature
of the neutron skin of $^{208}$Pb, and then the largest part of
$\Delta r_{np}$ is caused by $R_n \ne R_p$. We collect in
Table~\ref{TABLE2} the found equivalent sharp radii $R_n$ and $R_p$
and surface widths $b_n$ and $b_p$ of the densities of $^{208}$Pb in
the present mean-field models.

\begin{table}[t]
\caption{Equivalent sharp radius and surface width of the 2pF neutron
and proton density distributions of $^{208}$Pb in 
mean-field models. Units are fm.}
\begin{ruledtabular}
\begin{tabular}{lcccc}
Model   & $R_n$ & $R_p$ & $b_n$ & $b_p$ \\
\hline
HFB-8   & 6.822 & 6.782 & 0.991 & 0.819 \\
MSk7    & 6.847 & 6.808 & 0.980 & 0.801 \\
D1S     & 6.830 & 6.751 & 0.994 & 0.846 \\
SGII    & 6.890 & 6.806 & 0.971 & 0.821 \\
D1N     & 6.845 & 6.755 & 0.979 & 0.828 \\
Sk-T6   & 6.862 & 6.775 & 0.994 & 0.820 \\
HFB-17  & 6.883 & 6.797 & 0.996 & 0.821 \\
SLy4    & 6.902 & 6.790 & 1.007 & 0.852 \\
SkM*    & 6.907 & 6.786 & 1.007 & 0.847 \\
DD-ME2  & 6.926 & 6.800 & 1.026 & 0.829 \\
SkSM*   & 6.955 & 6.805 & 0.970 & 0.790 \\
SkMP    & 6.943 & 6.784 & 0.997 & 0.839 \\
FSUGold & 6.971 & 6.836 & 1.024 & 0.808 \\
Ska     & 6.970 & 6.789 & 0.998 & 0.844 \\
Sk-Rs   & 6.950 & 6.762 & 0.962 & 0.806 \\
Sk-T4   & 6.991 & 6.780 & 1.008 & 0.823 \\
G2      & 7.037 & 6.817 & 1.012 & 0.824 \\
NLC     & 7.087 & 6.863 & 1.016 & 0.820 \\
NL-SH   & 7.039 & 6.821 & 0.989 & 0.772 \\
TM1     & 7.085 & 6.862 & 1.005 & 0.787 \\
NL-RA1  & 7.065 & 6.834 & 1.008 & 0.797 \\
NL3     & 7.060 & 6.821 & 1.017 & 0.807 \\
NL3*    & 7.052 & 6.806 & 1.026 & 0.814 \\
NL-Z    & 7.134 & 6.865 & 1.058 & 0.847 \\
NL1     & 7.100 & 6.822 & 1.065 & 0.840
\end{tabular}
\end{ruledtabular}
\label{TABLE2}
\end{table}

\begin{figure}[t]
\includegraphics[width=1.0\columnwidth,clip=true]
{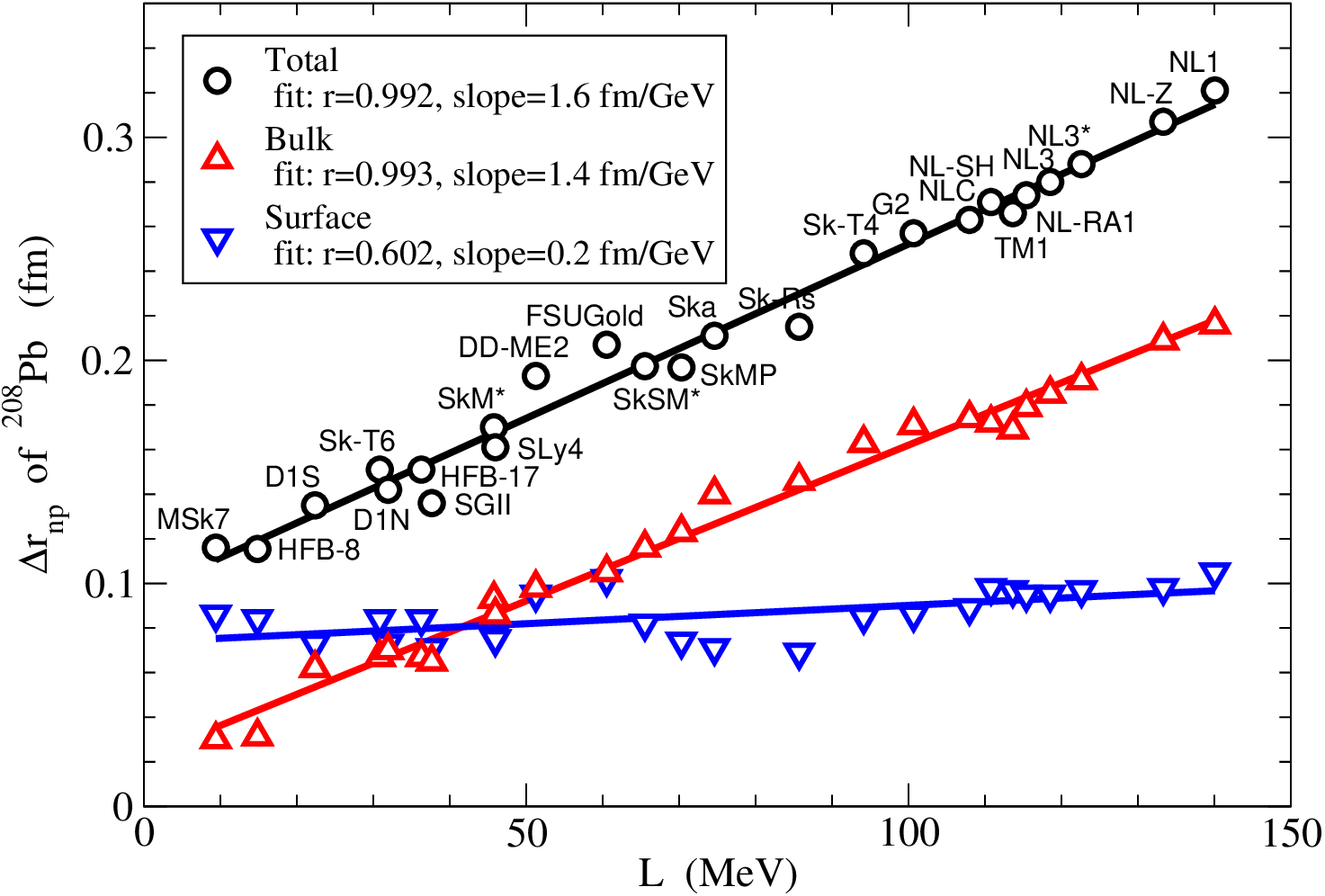}
\caption{\label{correlation}
(Color online) Correlation of the neutron skin thickness of
$^{208}$Pb and of its bulk and surface parts with the density
derivative $L$ of the nuclear symmetry energy.}
\end{figure}

As we have had the opportunity to see, the neutron skin thickness of a
heavy nucleus is strongly influenced by the density derivative $L$ of
the symmetry energy. Indeed, one easily suspects from
Table~\ref{TABLE1} that $\Delta r_{np}$ of $^{208}$Pb is almost
linearly correlated with $L$ in the nuclear mean-field models, which
Fig.~\ref{correlation} confirms for the present interactions. The
correlation of the neutron skin thickness of $^{208}$Pb with $L$ has
been amply discussed in the literature
\cite{fur02,dan03,bal04,ava07,cen09,cen09b,vid09}, as it implies that
an accurate measurement of the former observable could allow to
tightly constrain the density dependence of the nuclear symmetry energy. 
In particular, we studied the aforementioned correlation in Ref.\
\cite{cen09} where it is shown that the expression of the neutron skin
thickness in the DM of Myers and \'{S}wi\c{a}tecki \cite{MS,MSskin}
can be recast to leading order in terms of the $L$ parameter. To do
that, we use the fact that in all mean-field models the symmetry
energy coefficient computed at $\rho \approx 0.10$ fm$^{-3}$ is
approximately equal to the DM symmetry energy coefficient in
$^{208}$Pb which includes bulk- and surface-symmetry contributions
\cite{cen09}. In the standard DM, where the surface widths of the
neutron and proton densities are taken to be the same
\cite{MS,MSskin}, the neutron skin thickness is governed by the ratio
between the bulk-symmetry energy at saturation $J \equiv c_{\rm
sym}(\rho_0)$ and the surface stiffness coefficient $Q$ of the DM
\cite{cen09,cen09b} (the latter is not to be confused with the
equivalent rms radius $Q$ of Eq.\ (\ref{q})). The DM coefficient $Q$
measures the resistance of the nucleus against the separation of the
neutron and proton surfaces to form a neutron skin. We have shown
\cite{cen09,cen09b} in mean-field models that the DM formula for the
neutron skin thickness in the case where one assumes $b_n=b_p$,
undershoots the corresponding values computed by the semiclassical
extended Thomas-Fermi method in finite nuclei and, therefore, a
nonvanishing surface contribution is needed to describe more
accurately the mean-field results. However, this surface contribution
has a more involved dependence on the parameters of the interaction
and does not show a definite correlation with the $J/Q$ ratio (see
Fig.~4 of Ref.\ \cite{cen09b}).

Now, we wondered to which degree the correlation with $L$ of the
neutron skin thickness of $^{208}$Pb holds in its bulk and surface
parts extracted from actual mean-field densities. From our discussion
of the indications provided by the DM, we can expect this correlation
to be strong in the bulk part and weak in the surface part. Indeed,
the plots of $\Delta r_{np}^{\rm bulk}$ and $\Delta r_{np}^{\rm surf}$
against $L$ in Fig.~\ref{correlation} show that the bulk part displays
the same high correlation with $L$ as the total neutron skin (the
linear correlation factor is of 0.99 in both cases), whereas the
surface part exhibits a mostly flat trend with $L$. The linear fits in
Fig.~\ref{correlation} of the neutron skin thickness of $^{208}$Pb and
of its bulk part have also quite similar slopes. One thus concludes
that the linear correlation of $\Delta r_{np}$ of $^{208}$Pb with the
density content of the nuclear symmetry energy arises mainly from the
bulk part of $\Delta r_{np}$. In other words, the correlation arises
from the change induced by the density dependence of the symmetry
energy in the equivalent sharp radii of the nucleon density
distributions of $^{208}$Pb rather than from the change of the width
of the surface of the nucleon densities.

The value of about 0.1 fm that the surface contribution to
$\Delta r_{np}$ takes in $^{208}$Pb can be understood as follows
starting from Eq.\ (\ref{rs}). Taking into account that in 
2pF distributions fitted to mean-field densities $R_n \sim R_p \sim
1.16 A^{1/3}$ fm and $b_n + b_p \sim 1.8$ fm (see Table \ref{TABLE2}),
Eq.\ (\ref{rs}) can be approximated as
\begin{equation}
\Delta r^{\rm surf}_{np} \sim 3 A^{-1/3} (b_n -b_p) .
\end{equation}
Given that $b_n - b_p \sim 0.2$ fm for $^{208}$Pb on the average in
mean-field models (see Table \ref{TABLE2}), one finds $\Delta r^{\rm
surf}_{np} \sim$ 0.1 fm, rather independently of the model used to
compute it. It is interesting why the range of variation of $b_n$
with respect to $b_p$ is not larger in nuclear models, in view of the
fact that $R_n-R_p$ can take more different values. As discussed in
Ref.\ \cite{prex2}, this constraint is imposed on the models most
likely by the mass fits. For example, a model having nucleon densities
with very small or very large surface widths (i.e., very sharp or very
extended surfaces) would provoke a large change in the surface energy
of the nucleus, but that hardly would be successful to reproduce the
known nuclear masses.

\subsection{Discussion of the shape of the neutron density profiles}

The use of 2pF functions in order to represent the nuclear densities
by approximate distributions is also quite common in the experimental
investigations. The parameters of the proton 2pF distribution can be
assumed known in experiment, by unfolding from the accurately measured
charge density \cite{warda10}. 
However, the shape of the neutron density is more
uncertain, and even if the neutron rms radius is determined, it can
correspond to different shapes of the neutron density. Actually, the
shape of the neutron density is a significant question in the
extraction of nuclear information from experiments in exotic atoms
\cite{trz01,jas04,fried03,fried05} and from parity-violating
electron scattering \cite{don09}.
To handle the possible differences in the shape of the neutron density
when analyzing the experimental data, the so-called ``halo'' and
``skin'' forms are frequently used
\cite{trz01,jas04,fried03,fried05,don09}. In the ``halo-type''
distribution the nucleon 2pF shapes have $C_n = C_p$ and $a_n > a_p$,
whereas in the ``skin-type'' distribution they have $a_n = a_p$ and
$C_n > C_p$. To complete our study, we believe worth discussing the
predictions of the theoretical models for the parameters of the 2pF
shapes in $^{208}$Pb.

We compile in Table~\ref{TABLE3} the central radii $C_n$ and $C_p$ and
the diffuseness parameters $a_n$ and $a_p$ of the 2pF nucleon density
profiles of $^{208}$Pb obtained from the mean-field models of
Table~\ref{TABLE1}. We see that $C_n$ of neutrons spans a range of
approximately 6.7--6.85 fm in the nonrelativistic interactions and
that it is of approximately 6.8--7 fm in the relativistic parameter
sets. In the case of the proton density distribution, the value of
$C_p$ is smallest ($\sim$6.65 fm) in the two Gogny forces, it is about
6.67--6.71 fm in the Skyrme forces, and it is in a range of 6.7--6.77
fm in the RMF models. Then, we note that not only $C_n$ of neutrons
but also $C_p$ of protons is generally smaller in the nonrelativistic
forces than in the relativistic forces. The total spread in $C_p$
among the models (about 0.12 fm) is, though, less than half the spread
found in $C_n$ (about 0.3 fm). Indeed, the accurately known charge
radius of $^{208}$Pb is an observable that usually enters the fitting
protocol of the effective nuclear interactions.

\begin{table}[t]
\caption{Central radius and surface diffuseness of the 2pF neutron and
proton density distributions of $^{208}$Pb in 
mean-field models. Units are fm.}
\begin{ruledtabular}
\begin{tabular}{lcccccc}
Model   & $C_n$ & $C_p$ & $a_n$ & $a_p$ &$C_n-C_p$& $a_n-a_p$ \\
\hline
HFB-8   & 6.679 & 6.683 & 0.546 & 0.451 &$-$0.004 \ \ & 0.095 \\
MSk7    & 6.707 & 6.714 & 0.540 & 0.442 &$-$0.007 \ \ & 0.099 \\
D1S     & 6.687 & 6.649 & 0.546 & 0.464 &   0.038 & 0.082 \\
SGII    & 6.753 & 6.707 & 0.536 & 0.453 &   0.046 & 0.083 \\
D1N     & 6.705 & 6.657 & 0.537 & 0.453 &   0.048 & 0.084 \\
Sk-T6   & 6.718 & 6.676 & 0.548 & 0.452 &   0.042 & 0.096 \\
HFB-17  & 6.739 & 6.697 & 0.549 & 0.453 &   0.042 & 0.096 \\
SLy4    & 6.755 & 6.683 & 0.555 & 0.470 &   0.072 & 0.085 \\
SkM*    & 6.760 & 6.681 & 0.555 & 0.467 &   0.079 & 0.088 \\
DD-ME2  & 6.774 & 6.699 & 0.566 & 0.457 &   0.075 & 0.109 \\
SkSM*   & 6.819 & 6.713 & 0.535 & 0.436 &   0.106 & 0.099 \\
SkMP    & 6.799 & 6.680 & 0.550 & 0.463 &   0.119 & 0.087 \\
FSUGold & 6.821 & 6.740 & 0.564 & 0.446 &   0.081 & 0.118 \\
Ska     & 6.827 & 6.684 & 0.550 & 0.465 &   0.143 & 0.085 \\
Sk-Rs   & 6.817 & 6.665 & 0.530 & 0.444 &   0.152 & 0.086 \\
Sk-T4   & 6.846 & 6.681 & 0.555 & 0.453 &   0.165 & 0.102 \\
G2      & 6.891 & 6.717 & 0.558 & 0.454 &   0.174 & 0.104 \\
NLC     & 6.941 & 6.765 & 0.560 & 0.452 &   0.176 & 0.108 \\
NL-SH   & 6.900 & 6.733 & 0.546 & 0.426 &   0.167 & 0.120 \\
TM1     & 6.942 & 6.772 & 0.554 & 0.434 &   0.170 & 0.120 \\
NL-RA1  & 6.921 & 6.741 & 0.556 & 0.440 &   0.180 & 0.116 \\
NL3     & 6.914 & 6.726 & 0.560 & 0.445 &   0.188 & 0.115 \\
NL3*    & 6.903 & 6.709 & 0.566 & 0.449 &   0.194 & 0.117 \\
NL-Z    & 6.977 & 6.761 & 0.584 & 0.467 &   0.216 & 0.117 \\
NL1     & 6.940 & 6.718 & 0.587 & 0.463 &   0.222 & 0.124
\end{tabular}
\end{ruledtabular}
\label{TABLE3}
\end{table}

If we inspect the results for the surface diffuseness of the
density profiles of $^{208}$Pb in Table~\ref{TABLE3}, we see that $a_n$
of neutrons lies in a window of 0.53--0.59 fm (with the majority of
the models having $a_n$ between 0.545 and 0.565 fm). The
nonrelativistic interactions favor $a_n \lesssim 0.555$ fm, whereas
the RMF sets favor $a_n \gtrsim 0.555$ fm. This indicates that the
fall-off of the neutron density of $^{208}$Pb at the surface is in
general faster in the interactions with a soft symmetry energy than in
the interactions with a stiff symmetry energy. The surface diffuseness
$a_p$ of the proton density spans {\em in either} the nonrelativistic
or the relativistic models almost the same window of values
(0.43--0.47 fm; with the majority of the models having $a_p$ between 
0.445 and 0.465 fm). This fact is in contrast to the other 2pF
parameters discussed so far. Actually, the $a_p$ value of the proton
density can be definitely larger in some nonrelativistic forces than
in some relativistic forces (for example, in the case of SkM* and
NL3). One finds that the total spread of $a_n$ and $a_p$ within the
analyzed models is quite similar: about 0.05 fm in both $a_n$ and
$a_p$. This spread corresponds roughly to a 10\% variation compared to
the mean values of $a_n$ and $a_p$. It is remarkable that while among
the models $C_n$ has a significantly larger spread than $C_p$, the
surface diffuseness $a_n$ of the neutron density has essentially the
same small spread as the surface diffuseness $a_p$ of the proton
density. As we have discussed at the end of Sec.\ \ref{bulksurf}, this
is likely imposed by the nuclear mass fits. It means that our
ignorance about the neutron distribution in $^{208}$Pb does not seem
to produce in the mean-field models a larger uncertainty for $a_n$ of
neutrons than for $a_p$ of protons, and that most of the uncertainty
goes to the value of $C_n$.

\begin{figure}
\includegraphics[width=0.98\columnwidth,clip=true]
{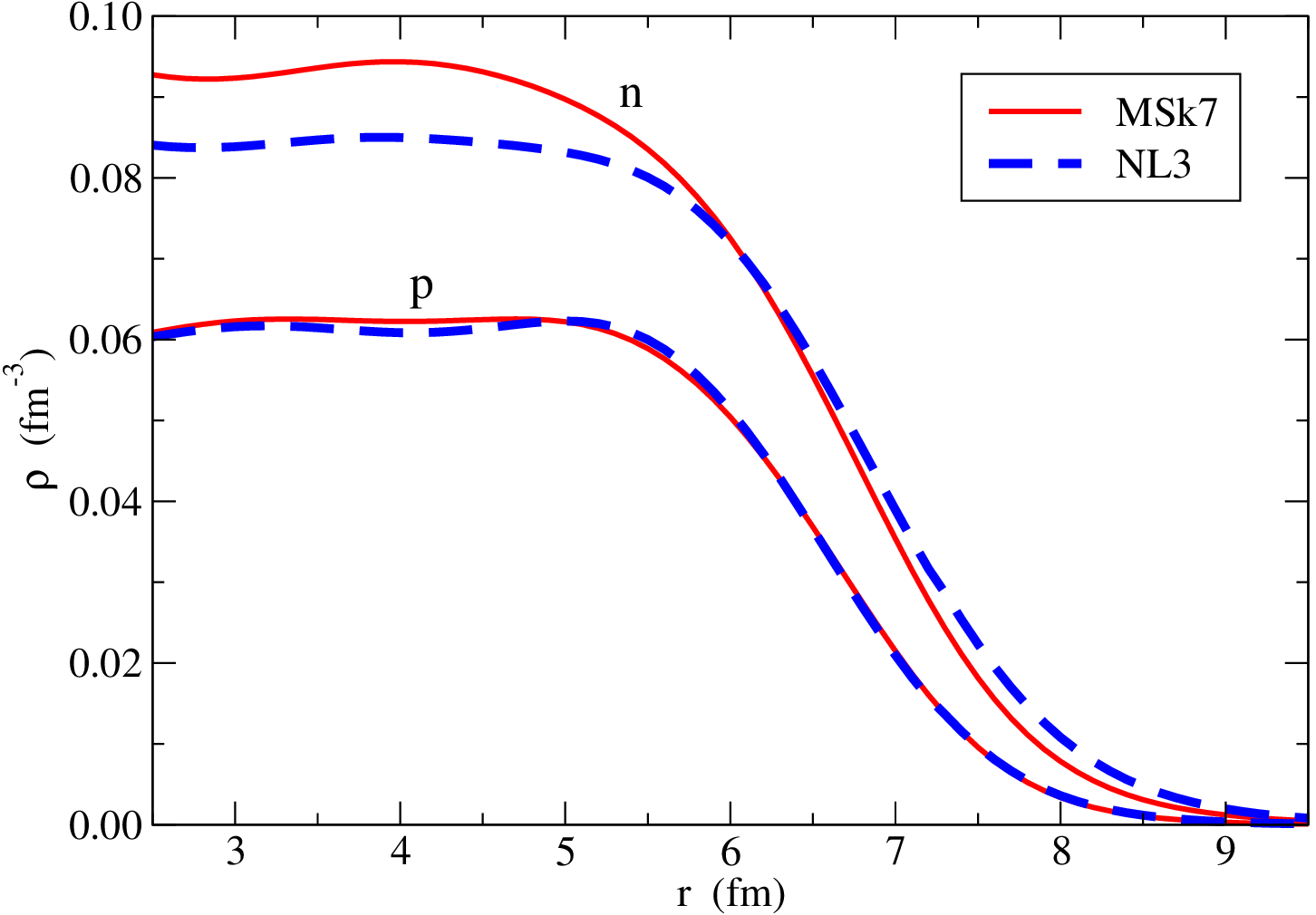}
\caption{\label{profiles}
(Color online)
Comparison of the nucleon densities predicted in $^{208}$Pb by the
mean-field models MSk7 ($C_n-C_p \approx 0$ fm) and NL3 ($C_n-C_p
\approx 0.2$ fm).}
\end{figure}

The difference $C_n-C_p$ of the central radii of the nucleon densities
of $^{208}$Pb turns out to range approximately between 0.~and 0.2 fm.
It is smaller for soft symmetry energies and larger for stiff symmetry
energies. We realize that the limiting situation of a halo-type
distribution where the nucleon densities of $^{208}$Pb have $C_n =
C_p$ and $a_n>a_p$ is actually attained in the nuclear mean-field
models with a very soft symmetry energy (like in HFB-8 or MSk7 where
$C_n-C_p$ even is slightly negative). 
The difference $a_n-a_p$ of the neutron and proton surface diffuseness
in $^{208}$Pb is comprised between nearly 0.08 and 0.1 fm in the
nonrelativistic forces and between nearly 0.1 and 0.12 fm in the RMF
forces. This implies that no interaction predicts $a_n-a_p$ of
$^{208}$Pb as close to vanishing as $C_n-C_p$ is in some forces. Thus,
the limiting situation where the nucleon densities in $^{208}$Pb would
have $a_n=a_p$ and $C_n > C_p$ is not found in the nuclear mean-field
models. Indeed, we observe in Table~\ref{TABLE3} that if $C_n-C_p$
becomes larger in the models, also $a_n-a_p$ tends overall to become
larger. 
In order to help visualize graphically the change in the mean-field
nucleon densities of $^{208}$Pb from having a nearly vanishing $C_n-C_p$
or a large $C_n-C_p$, we have plotted in Fig.~\ref{profiles} the
example of the densities of the MSk7 and NL3 interactions.
On the one hand, we see that both models MSk7 and NL3 predict
basically the same proton density, as expected. On the other hand, the
difference between having $C_n \approx C_p$ in MSk7 and $C_n > C_p$ in
NL3 can be appreciated in the higher bulk and the faster fall-off at
the surface of the neutron density of MSk7 compared with NL3.

In summary, we conclude that in $^{208}$Pb the nuclear mean-field
models favor the halo-type distribution with $C_n\approx C_p$ and $a_n
> a_p$ if they have a very soft (``supersoft'') symmetry energy, they
favor a mixed-type distribution if they have mild symmetry energies,
and a situation where $C_n$ is clearly larger than $C_p$ if the
symmetry energy is stiff, but that the pure skin-type distribution where
$a_n-a_p=0$ in $^{208}$Pb is not supported (not even $a_n-a_p \approx
0$) by the mean-field models. Although the experimental evidence
available to date on the neutron skin thickness of $^{208}$Pb is
compatible with the ranges of the $C_n-C_p$ and $a_n-a_p$ parameters
considered in our study, it is not to be excluded that the description
of a precision measurement in $^{208}$Pb may need of nucleon densities
with $C_n-C_p$ or $a_n-a_p$ values not fitting Table~\ref{TABLE3}.
However, a sizable deviation (such as $a_n-a_p=0$) could mean that
there is some missing physics in the isospin channel of present
mean-field interactions, because once these interactions are
calibrated to reproduce the observed binding energies and charge radii
of nuclei they typically lead to the ranges of Table~\ref{TABLE3}.

\subsection{Application to parity-violating electron scattering}


Parity-violating electron scattering is expected to be able to
accurately determine the neutron density in a nucleus since the $Z^0$
boson couples mainly to neutrons \cite{prex1,prex2}. Specifically, the
PREX experiment \cite{prex1} aims to provide a clean measurement of
the neutron radius of $^{208}$Pb. In this type of experiments one
measures the parity-violating asymmetry
\begin{equation}
A_{LR}\equiv
\frac{\displaystyle \frac{d\sigma_+}{d\Omega}-\frac{d\sigma_-}{d\Omega}}
{\displaystyle \frac{d\sigma_+}{d\Omega}+\frac{d\sigma_-}{d\Omega}} ,
\label{alr}
\end{equation}
where $d\sigma_\pm/d\Omega$ is the elastic electron-nucleus cross
section. The plus (minus) sign accounts for the fact that electrons
with a positive (negative) helicity state scatter from different
potentials ($V_\pm(r)=V_{\rm Coulomb}(r) \pm V_{\rm weak}(r)$ for
ultra-relativistic electrons). 
Assuming for simplicity the plane wave Born approximation (PWBA) and
neglecting nucleon form factors, the parity-violating asymmetry at
momentum transfer $q$ can be written as \cite{prex2}
\begin{equation}
A_{LR}^{\rm PWBA} = \frac{G_F q^2}{4\pi \alpha \sqrt{2}}
\Big[ 4 \sin^2\theta_W + \frac{F_n(q) - F_p(q)}{F_p(q)} \Big] ,
\label{alrPWBA}
\end{equation}
where $\sin^2\theta_W \approx 0.23$ for the Weinberg angle and
$F_n(q)$ and $F_p(q)$ are the form factors of the point neutron and
proton densities. Because $F_p(q)$ is known from elastic electron
scattering, it is clear from (\ref{alrPWBA}) that the largest
uncertainty to compute $A_{LR}$ comes from our lack of knowledge of
the distribution of neutrons inside the nucleus. PREX intends to
measure $A_{LR}$ in $^{208}$Pb with a 3\% error (or smaller). This
accuracy is thought to be enough to determine the neutron rms radius
with a 1\% error \cite{prex1,prex2}.

\begin{figure}[t]
\includegraphics[width=0.98\columnwidth,clip=true]
{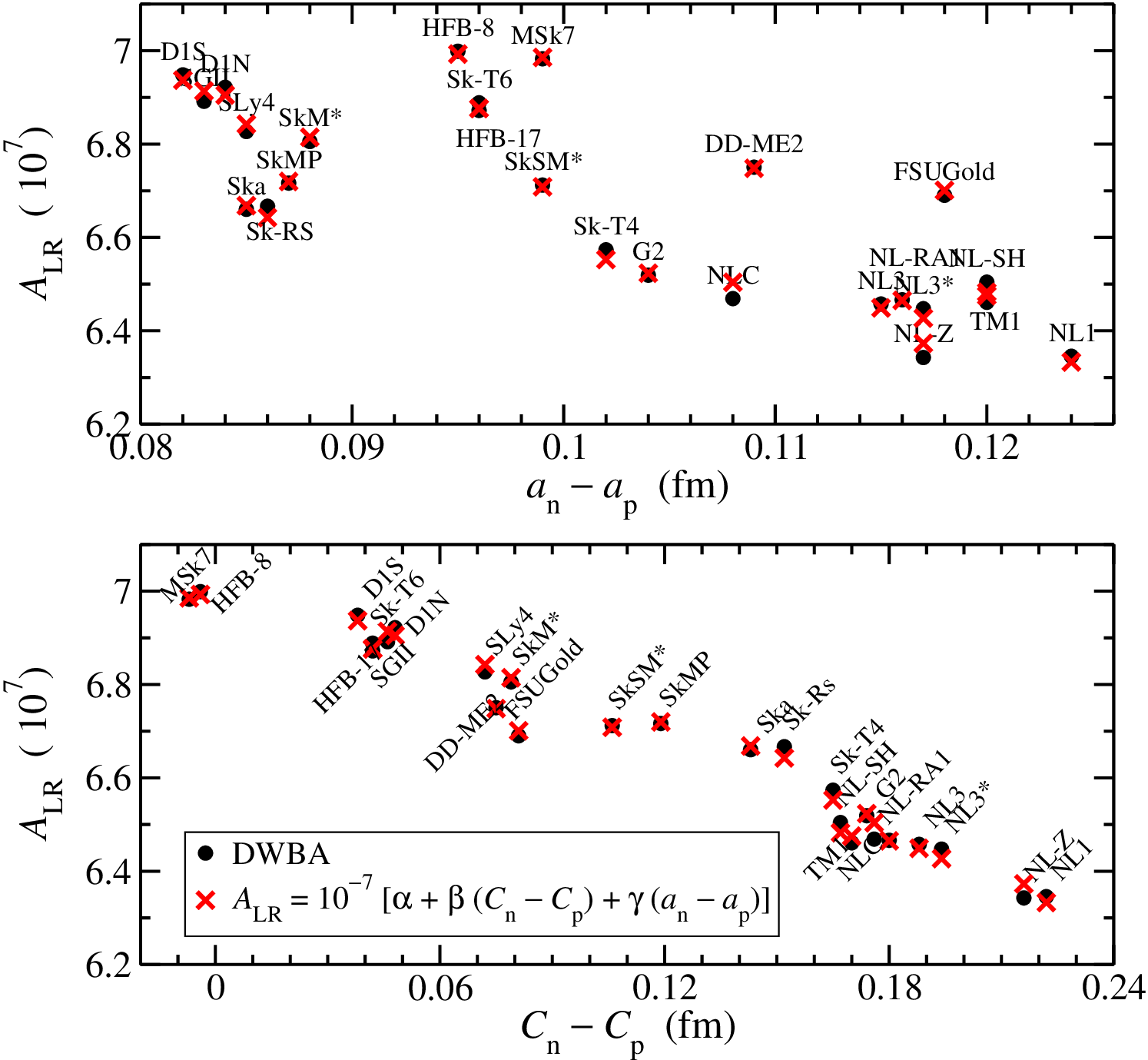}
\caption{\label{alr2pF} 
(Color online) Parity-violating asymmetry for 1~GeV electrons at
$5^\circ$ scattering angle calculated from the 2pF neutron and proton 
density distributions of $^{208}$Pb in nuclear mean-field models.}
\end{figure}

To compute the parity-violating asymmetry we essentially follow
the procedure described in Ref.\ \cite{prex2}. For realistic results,
we perform the exact phase shift analysis of the Dirac equation for
electrons moving in the potentials $V_\pm(r)$ \cite{roca08}. This
method corresponds to the distorted wave Born approximation (DWBA).
The main input needed for solving this problem are the charge and weak
distributions. To calculate the charge distribution, we fold the
mean-field proton and neutron point-like densities with the
electromagnetic form factors provided in \cite{emff}. For the weak
distribution, we fold the nucleon point-like densities with the
electric form factors reported in \cite{prex2} for the coupling of a
$Z^0$ to the protons and neutrons. We neglect the strange form factor
contributions to the weak density \cite{prex2}. Because the
experimental analysis may involve parametrized densities, in our study
we use the 2pF functions extracted from the self-consistent
densities of the various models. The difference between $A_{LR}$
calculated in $^{208}$Pb with the 2pF densities and with the
self-consistent densities is anyway marginal at most. In
Fig.~\ref{alr2pF}, $A_{LR}^{\rm DWBA}$ obtained with the Fermi
distributions listed in Table \ref{TABLE3} is plotted against the
values of $C_n-C_p$ (lower panel) and $a_n-a_p$ (upper panel). To
simulate the kinematics of the PREX experiment \cite{prex1}, we set
the electron beam energy to 1~GeV and the scattering angle to
$5^\circ$, which corresponds to a momentum transfer in the laboratory
frame of $q=0.44$ fm$^{-1}$.

First, one can see from Fig.~\ref{alr2pF} that the mean-field
calculations constrain in a rather narrow window the value of the
parity-violating asymmetry in $^{208}$Pb. The increasing trend of
$A_{LR}$ with decreasing $C_n-C_p$ indicates that $A_{LR}$ is larger
when the symmetry energy is softer. Note that a
large value $A_{LR} \approx 7 \times 10^{-7}$ (at 1~GeV and $5^\circ$)
would be in support of a more surface than bulk origin of the neutron
skin thickness of $^{208}$Pb and of the halo-type density distribution
for this nucleus. Second, $A_{LR}^{\rm DWBA}$ displays in good
approximation a linear correlation with $C_n-C_p$ ($r=0.978$), while
the correlation with $a_n-a_p$ is not remarkable.
Nevertheless, we have found a very good description of $A_{LR}^{\rm
DWBA}$ of the mean-field models---well below the 3\% limit of
accuracy of the PREX experiment---by means of a fit in $C_n-C_p$ and
$a_n-a_p$ (red crosses in Fig.~\ref{alr2pF}):
\begin{equation}
A_{LR}^{\rm fit} = [ \alpha
 +  \beta (C_n-C_p) +  \gamma (a_n-a_p) ]\times 10^{-7} ,
\label{alrfit}
\end{equation}
with $\alpha=7.33$, $\beta=-2.45$ fm$^{-1}$, and $\gamma=-3.62$ fm$^{-1}$. 

The parametrization (\ref{alrfit}) may be easily
understood if we consider the PWBA expression
of $A_{LR}$ given above in Eq.\ (\ref{alrPWBA}).
At low-momentum transfer, the form factors $F_n(q)$ and $F_p(q)$ of
the neutron and proton densities (these are point densities in PWBA)
can be expanded to first order in $q^2$, so that the numerator inside
brackets in Eq.\ (\ref{alrPWBA}) becomes
$-(q^2/6)\, (\langle r^2 \rangle_n - \langle r^2 \rangle_p)$.
In 2pF density distributions we have
$\langle r^2 \rangle_q = (3/5)\, C_q^2 + (7\pi^2/5)\, a_q^2$. Now,
assuming constancy of $F_p(q^2)$ in the nuclear models and taking into
account that $C_n+C_p \gg C_n-C_p$ and $a_n+a_p \gg a_n-a_p$, it is
reasonable to assume that the variation of $A_{LR}$ is dominated by
the change of $C_n-C_p$ and $a_n-a_p$ as proposed in Eq.~(\ref{alrfit}).

In the analysis of a measurement of $A_{LR}$ in $^{208}$Pb through
parametrized Fermi densities, one could set $C_p$ and $a_p$ to those
known from experiment \cite{warda10} and then vary $C_n$ and $a_n$ in
(\ref{alrfit}) to match the measured value. According to the
predictions of the models in Table~\ref{TABLE3}, it would be
reasonable to restrict this search to windows of about 0--0.22 fm for
$C_n-C_p$ and 0.08--0.125 fm for $a_n-a_p$. Therefore, the result of a
measurement of the parity-violating asymmetry together with
Eq.~(\ref{alrfit}) (or Fig.~\ref{alr2pF}) would allow not only to
estimate the neutron rms radius of $^{208}$Pb but also to obtain some
insight about the neutron density profile in this nucleus. This
assumes that the experimental value for $A_{LR}$ will fall in, or at
least will be not far from, the region allowed by the mean-field
calculations at the same kinematics.


\section{Summary}
\label{summary}


We have investigated using Skyrme, Gogny, and relativistic mean-field
models of nuclear structure whether the difference between the
peripheral neutron and proton densities that gives rise to the neutron
skin thickness of $^{208}$Pb is due to an enlarged bulk radius of
neutrons with respect to that of protons or, rather, to the difference
between the widths of the neutron and proton surfaces. The
decomposition of the neutron skin thickness in bulk and surface
components has been obtained through two-parameter Fermi distributions
fitted to the self-consistent nucleon densities of the models.

Nuclear models that correspond to a soft symmetry energy, like various
nonrelativistic mean-field models, favor the situation where the size
of the neutron skin thickness of $^{208}$Pb is divided similarly into
bulk and surface components. If the symmetry energy of the model
is ``supersoft'', the surface part even becomes dominant.
Instead, nuclear models that correspond to a stiff symmetry energy,
like most of the relativistic models, predict a bulk component about
twice as large as the surface component. We have found that the size
of the surface component changes little among the various nuclear
mean-field models and that the known linear correlation of $\Delta
r_{np}$ of $^{208}$Pb with the density derivative of the nuclear
symmetry energy arises from the bulk part of $\Delta r_{np}$. The
latter result implies that an experimental determination of the
equivalent sharp radius of the neutron density of $^{208}$Pb could be
as useful for the purpose of constraining the density-dependent
nuclear symmetry energy as a determination of the neutron rms radius.

We have discussed the shapes of the 2pF distributions predicted for
$^{208}$Pb by the nuclear mean-field models in terms of the so-called
``halo-type'' ($C_n-C_p=0$) and ``skin-type'' ($a_n-a_p=0$)
distributions of frequent use in experiment. It turns out that the
theoretical models can accomodate the halo-type distribution in
$^{208}$Pb if the symmetry energy is supersoft. However, they do no
support a purely skin-type distribution in this nucleus, even if the
model has a largely stiff symmetry energy. Let us mention that the
information on neutron densities from antiprotonic atoms favored the
halo-type over the skin-type distribution \cite{trz01,jas04}.

We have closed our study with a calculation of the asymmetry $A_{LR}$
for parity-violating electron scattering off $^{208}$Pb in conditions
as in the recently run PREX experiment \cite{prex1}, using the
equivalent 2pF shapes of the models. This has allowed us to find a
simple parametrization of $A_{LR}$ in terms of the differences
$C_n-C_p$ and $a_n-a_p$ of the parameters of the nucleon
distributions. With a measured value of the parity-violating
asymmetry, it would provide a new correlation between the central
radius and the surface diffuseness of the distribution of neutrons in
$^{208}$Pb, assuming the same properties of the proton density known
from experiment.

\begin{acknowledgments}
Work partially supported by the Spanish Consolider-Ingenio 2010
Programme CPAN CSD2007-00042 and by Grants No.\ FIS2008-01661 from
MICIN (Spain) and FEDER, No.\ 2009SGR-1289 from Generalitat de
Catalunya (Spain), and No.\ N202~231137 from MNiSW (Poland).
\end{acknowledgments}



\end{document}